\documentclass[11pt]{article}
\pdfoutput=1

\usepackage[utf8]{inputenc}
\usepackage{multirow}
\usepackage{amsmath, amsfonts, amssymb}
\usepackage{comment}
\usepackage{graphicx}
\usepackage{psfrag}
\usepackage{amsthm}
\usepackage[usenames,dvipsnames,svgnames,table]{xcolor}
\usepackage{enumerate}
\usepackage{arydshln}
\usepackage{soul}
 \usepackage{slashed}
\usepackage{subfig}
\usepackage{mathrsfs}
 \usepackage{a4wide}
  \usepackage{tikz}
  \usepackage{tikz-cd}
  \usetikzlibrary{shapes.geometric}
  \usepackage{tcolorbox}

  \usepackage{color}
  \definecolor{dark-gray}{gray}{0.20}
  \definecolor{gray}{gray}{0.30}
  \definecolor{light-gray}{gray}{0.80}
  \definecolor{dark-red}{rgb}{0.7,0,0}
  \definecolor{dark-green}{rgb}{0.1,0.4,0}
  \definecolor{dark-blue}{rgb}{0.3,0.3,0.7}
  \definecolor{light-blue}{rgb}{0.8,0.8,1}
      \definecolor{swamp}{RGB}{240, 199, 197}
       \definecolor{landscape}{RGB}{180, 250, 199}
          \definecolor{undecided}{RGB}{252, 252, 197}

  \usepackage{pifont}

\usepackage{newunicodechar} 

\usepackage{setspace}

\newcommand{\be}{\begin{equation}}
\newcommand{\ee}{\end{equation}}
\newcommand{\eq}[1]{(\ref{#1})}

\def\be{\begin{equation}}
\def\ee{\end{equation}}
\def\bea{\begin{eqnarray}}
\def\eea{\end{eqnarray}}

\def\simleq{\; \raise0.3ex\hbox{$<$\kern-0.75em
      \raise-1.1ex\hbox{$\sim$}}\; }
   \def\simgeq{\; \raise0.3ex\hbox{$>$\kern-0.75em
      \raise-1.1ex\hbox{$\sim$}}\; }

\numberwithin{equation}{section}

\usepackage{jheppub}
\usepackage{hyperref}
\usepackage{cleveref}

\hypersetup{
	colorlinks=true,
	linkcolor=dark-blue,
	citecolor=dark-red,
	urlcolor=dark-green,
	linktoc=page
}

\theoremstyle{remark}

\crefname{appendix}{Appendix}{Appendices}

\title{\centering Pure Supersymmetric $AdS$ and the Swampland}

\author{Miguel Montero$^1$,}\affiliation{$^1$Department of Physics, Harvard University, Cambridge, MA 02138, USA}
\author{Martin Ro\v{c}ek$^2$,} \affiliation{$^2$C.N.Yang Institute for Theoretical Physics, Stony Brook University,
Stony Brook, NY 11794-3840,USA}
\author{Cumrun Vafa$^1$} 

\emailAdd{mmontero@g.harvard.edu}
\emailAdd{martin.rocek@stonybrook.edu}
\emailAdd{vafa@g.harvard.edu}

\abstract{We point out that pure supergravity theories in $AdS$ with enough supersymmetry lead, upon taking the large radius limit, to flat space quantum gravities with a nonperturbatively exact global symmetry, and are therefore in the Swampland. The argument applies to any $AdS$ supergravity with gauged R-symmetry group, including truncations of most well known examples, such as $AdS_5$ without the $S^5$ or $AdS_4$ without the $S^7$.
 This demonstrates that extreme scale separation, at least with enough supersymmetry, is not realizable. Moreover pure $AdS$ theories are also in conflict with some other Swampland principles including the Weak Gravity Conjecture and the (generalized) Distance Conjecture.  }

\begin{document}
\hypersetup{pageanchor=false}
\makeatletter
\let\old@fpheader\@fpheader

\makeatother

\maketitle

\hypersetup{
    pdftitle={Pure $AdS$ and the Swampland},
    pdfauthor={ Miguel Montero, Martin Rocek, Cumrun Vafa},
    pdfsubject={Swampland and $AdS$}
}

\newcommand{\remove}[1]{\textcolor{red}{\sout{#1}}}

\section{Introduction}
The $AdS$/CFT correspondence (see e.g. \cite{Zaffaroni:2000vh,Hubeny:2014bla} for reviews) provides a non-perturbative definition of what we mean by quantum gravity in $AdS$ space. Although the idea behind the correspondence is logically independent from string theory, all known examples so far where the $AdS$ gravity is Einsteinian arise from string theory constructions, where the CFT corresponds to the worldvolume theory of a certain stack of branes probing a geometry, and the $AdS$ vacuum describes its near-horizon geometry. 

A very important quantity in any holographic pair is the size of the internal dimensions relative to the curvature scale of the $AdS$ space. A model where the extra dimensions are much smaller than the $AdS$ scale is said to be ``scale-separated''.  Let $m$ denote the KK mass scale.  In general one expects
\begin{equation}
m\sim |\Lambda |^a \rightarrow R\sim \ell^{2a}
\end{equation}
where $a\geq 0$, $\ell$ is the $AdS$ length, and $R$ is the radius of the internal dimension.  The case $a=\frac12$ occurs when the internal dimension is of the same scale as the $AdS$ scale. For $0<a<\frac12$ the internal dimension will be smaller, but will still grow with the $AdS$ scale.  The extreme case of scale separation occurs when $R$ is fixed and independent of $\ell$; for example it can be Planckian.  If achievable, scale separation could serve as a promising ingredient in proposed constructions of de Sitter vacua \cite{Kachru:2003aw,Balasubramanian:2005zx} -- if they indeed exist.

Given its importance, it is remarkable that no example with an explicit embedding in string theory is scale separated. For instance, in famous examples such as as IIB-string theory on $AdS_5\times S^5$ or $AdS_3\times S^3\times T^4$, and M-theory on $AdS_7\times S^4$ or $AdS_4\times S^7$, the sphere factors are of the same size as the $AdS$, which corresponds to $a=\frac12$.

The strong form of the $AdS$ Distance Conjecture proposes that for the supersymmetric case $a=\frac12$ \cite{Lust:2019zwm}. There are proposed supersymmetric constructions with scale separation which are consistent only with the weak form of the $AdS$ Distance conjecture, which has $a<\frac12$, such as DGKT \cite{DeWolfe:2005uu}, but their validity is still a debated topic of active research \cite{Alday:2019qrf,Junghans:2018gdb,Banlaki:2018ayh,VanHemelryck:2022ynr,Junghans:2020acz,Cribiori:2021djm,Conlon:2020wmc,Farakos:2020phe,Marchesano:2022rpr,Marchesano:2021ycx,Quirant:2022fpn,Marchesano:2020uqz,Marchesano:2020qvg,Marchesano:2019hfb,Apers:2022vfp,Apers:2022tfm,Apers:2022zjx}.

In this note, we will consider an extreme version of scale separation, corresponding to $a=0$, which is to assume that the internal dimensions are not there at all or are frozen at the Planck scale: one just has a pure $d$-dimensional theory of gravity, with vacuum $AdS_d$, dual to CFT$_{d-1}$. A particular case of this are theories including fields in the gravity (super)multiplet only -- theories of ``pure'' supergravity. The non-supersymmetric version of this question (the existence of pure gravity in $AdS$) remains open after much work, even for $AdS_3/CFT_2$. Recently, \cite{Alday:2022ldo} considered the holographic dual to pure $\mathcal{N}=8$ SUGRA\footnote{We are counting all supercharges, corresponding to bulk killing spinors;  
so for instance, $AdS_5\times S^5$ has $(\mathcal{N}=8)\cdot 4=32$ real supercharges. } in $AdS_5$ in the large $AdS$ limit, which is dual to the $\mathcal{N}=4$ stress-energy tensor superconformal multiplet (and multi-trace descendants of it); they found that the model satisfied, and in fact saturated, certain bootstrap constraints. This was interpreted in \cite{Alday:2022ldo} as indirect evidence that pure $\mathcal{N}=8$ SUGRA in $AdS_5$, without the $S^5$ or any other internal dimensions, might make sense as a consistent quantum gravity.

We also notice that, with enough supersymmetry, pure supergravity in Minkwoski (without a reference to AdS), regarded as an effective field theory valid up to any given energy scale $E$, is also in the Swampland, simply because we can go to a point in moduli space where BPS black holes become arbitrarily light, thanks to the Swampland Distance Conjecture, even if the effective field theory of the massless sector always remains the same. 

The purpose of this short note is to present a simple argument that with enough supersymmetry, including pure $\mathcal{N}=8$ SUGRA in $AdS_5$, in the infinite radius limit ( the ``large $N$'' limit) cannot be a consistent quantum gravitational theory. The reason is that any of these theories defines in this limit  a flat-space quantum gravity with a global symmetry. Absence of global symmetries is perhaps the best established Swampland constraint, and is amply supported by several independent arguments \cite{Banks:1988yz,Banks:2010zn,Harlow:2018tng,Chen:2020ojn} (see \cite{vanBeest:2021lhn} for a review). Additionally, the theories thus constructed are also in trouble with the Distance Conjecture as well as the Weak Gravity Conjecture and the Completeness Hypothesis. In fact, the connection to the lattice/tower WGC conjectures has been already pointed out (in the abelian case) in \cite{Cribiori:2022trc} (the argument there was phrased in terms of the magnetic WGC cutoff, which is a true EFT cutoff only if there is a tower of WGC particles). Indeed, the WGC tower of states is often the way that global symmetries become obstructed in quantum gravity \cite{Gendler:2020dfp}. Although we agree with the conclusions of \cite{Cribiori:2022trc}, direct application of magnetic/lattice WGC to $AdS$ is a strong assumption, particularly since lattice WGC is not always true even in perturbative limits \cite{Heidenreich:2016aqi}, where only a sublattice of superextremal states exists, which in principle can be of arbitrary index. Furthermore, the distinction between single and multi-particle states is murky in $AdS$, since it behaves as a box, and there is an obvious tower of superextremal multi-particle  states -- multi-particle states of the $SU(4)$ vector bosons themselves. Finally, the extremality condition, and the meaning of WGC in $AdS$, are likely to be different from their flat space counterparts (see e.g. \cite{Nakayama:2015hga}).  The  point of this note is to sidestep these issues and directly spell out what exactly is bad if the tower of states is absent: one gets to a quantum gravity with an exact global symmetry.

We will illustrate the argument in detail with the case of pure $\mathcal{N}=8$ five-dimensional supergravity in Section \ref{sec:5dsugra}; in Section \ref{sec:list}, we collect several other examples where a similar argument applies. Finally, Section \ref{sec:conclu} contains a few concluding remarks.

\section{Pure \texorpdfstring{$\mathcal{N}=8$}{N=8} SUGRA in five dimensions and extreme scale separation}\label{sec:5dsugra}
Consider the standard $AdS_5\times S^5$ solution of IIB string theory, dual to $\mathcal{N}=4$ SYM. The $S^5$ is supported by $N$ units of RR flux threading it; isometries of $S^5$ appear as a $Spin(6)\approx SU(4)$ symmetry, which is gauged. The radius of the $S^5$ is equal to the $AdS$ lengthscale, $\ell$. This flux also controls the hierarchy between the five-dimensional Planck length\footnote{Defined by an Einstein term $\frac{1}{2\ell_5^3}\int d^5x\, R$.}, $\ell_5$, and the $AdS$ lengthscale $\ell$:
\begin{equation}\left(\frac{\ell}{\ell_5}\right)^3\propto N^2,\label{2ptt}\end{equation}
both of which appear in the five-dimensional lagrangian. Classically, there is a consistent truncation of the full IIB supergravity, described by a five-dimensional gauged supergravity \cite{Pernici:1985ju,Gunaydin:1984qu}, with the following field content :

\begin{itemize}
\item A real scalar in the $\mathbf{20}'$ (vector symmetric traceless) of $SU(4)$ of mass $m^2\ell^2=-4$.  This is dual to a scalar of conformal dimension $\Delta=2$.
\item Complex scalars in the $\mathbf{10}$ and $\mathbf{1}$ of $SU(4)$, of masses $m^2\ell^2=-3,0$ respectively. Dual to scalars of dimensions $\Delta=3,0$ respectively.
\item Vectors in the $\mathbf{15}$ (adjoint (of $SU(4)$), which are massless. These are $SU(4)$ gauge bosons, dual to the corresponding conserved currents of dimension $\Delta=3$ in the dual field theory.
\item A two-form $B_{\mu\nu}$ in the $\mathbf{6}$ of $SU(4)$, plus its complex conjugate. This is dual to a two-form in the dual field theory, of dimension $\Delta=3$. 
\item The graviton $g_{\mu\nu}$, dual to the stress-energy tensor of the dual CFT
\item Superpartners: Gravitini in the $\mathbf{4}$ of $SU(4)$ and fermions in the $\mathbf{4}$ and $\mathbf{10}$.
\end{itemize}
 The classical truncation described above is dual to the stress-energy tensor short superconformal multiplet \cite{Dolan:2002zh}, whose primary operator is a single-trace scalar in the $\mathbf{20'}$ built as $\text{Tr}(\Phi^{(i} \Phi^{j)})$ in terms of fundamental SYM fields. 
 
 With this much supersymmetry, the lagrangian and interactions are completely determined. In particular, the gauge coupling of the $SU(4)$ R-symmetry vectors (which is dimensionful) is related to the two-point function of the dual current, and it also scales as
 \begin{equation} \frac{\ell}{g^2}\sim N^2.\end{equation}
 Using \eq{2ptt}, we can write down the $SU(4)$ gauge coupling in five-dimensional Planck units, as
 \begin{equation} \frac{\ell_5}{g^2}\sim N^{\frac43},\label{key}\end{equation}
 where $g$ is the gauge coupling of the bulk dual $U(1)$ current.
 Focus now on the physics at a given energy $E$ which we keep fixed in five-dimensional Planck units. 
Equation \eq{key} tells us that, as we take the large $N$ limit, the five-dimensional gauge coupling goes to zero. At the same time, the size of the $S^5$ grows  with $N$, so that a tower of KK states becomes light in Planck units at a rate given by $m_{KK}\sim N^{-\frac23}$. The corresponding species scale, 
 goes to zero as $N\rightarrow\infty$, signalling that the infinite $N$ limit cannot be regarded as a five-dimensional gravity theory. Of course, the physics of this limit is very simple: As $N\rightarrow\infty$ the $S^5$ decompactifies and we land in a ten-dimensional theory of gravity, which is simply IIB.  In fact, this limiting procedure, known as the ``bulk-point limit'' in the holographic literature (see e.g. \cite{Lam:2017ofc,Giddings:1999jq}, has been used to study type IIB string theory in ten dimensions.

Imagine that we now try to do the same in a putative large $N$ family of five-dimensional pure supergravities, defined as a quantum gravity in $AdS$ where the only light fields are those of the gravity multiplet described above, and then there is nothing else until energies of the order the five-dimensional Planck scale, where we hit the black hole threshold.  Here by ``$N$'' we simply mean a parameter fixing the scale of $AdS$ as in the $AdS_5\times S^5$ case.  In the dual field theory, the spectrum of single-trace operators only contains those in the stress-energy tensor multiplet, and then there is a large gap until the Planck scale.

In the $N\rightarrow\infty$ limit, we must recover a five dimensional theory of gravity with 32 supercharges and vanishing cosmological constant -- an ungauged supergravity. An important feature of this theory is that it preserves the $SU(4)$ R-symmetry exactly, since at any finite $N$ the symmetry is gauged and is selection rules are exactly preserved. However, in the infinite $N$ limit, \eq{key} suggests that this symmetry is becoming global, and since there is no tower of KK modes to obstruct the infinite distance limit, the theory posesses a global symmetry at finite Planck mass, which is not acceptable.  

We can be more precise, due to the supersymmetry. There is only one five-dimensional supergravity with 32 supercharges and a Minkowski vacuum, described effective field theory, which can be described as the low-energy limit of either M-theory on $T^6$ or type II on $T^5$.  One can check that the $AdS_5$ fields described above match, in the infinite $N$ limit, the fields of this supergravity. For instance, M-theory on $T^6$ has 42 massless scalars, coming from deformations of $T^6$ (which gives 20 scalars), periods of the $C_3$ on 3-cycles (which give $\binom{6}{3}=20$) scalars, and the single scalar coming from the period of $C_6$ on the $T^6$, matching the 42 real scalars above whose masses go to zero in Planck units. One also has $\binom{6}{2}=15$ vectors coming from the periods of $C_3$ on 2-cycles of $T^6$, six 2-forms from $C_3$ along a 1-cycle of $T^6$, as well as six KK vectors that can be dualized to 2-forms in five dimensions. Together with the metric, all these fields provide a perfect match. Yet there is a fundamental difference: the vectors in the flat space theory are all abelian, so they cannot be gauge bosons for an $SU(4)$ symmetry. This yet again shows the symmetry must be global. Indeed, in the realization of this supergravity from string theory, the R-symmetry is always explicitly broken to a discrete subgroup by massive charged states. In fact, this supergravity has a famous $E_{6(6)}$ non-compact global symmetry group \cite{Cremmer:1979uq}, which is broken to a discrete duality subgroup in the usual embedding in M-theory. The $SU(4)$ global symmetry we have obtained in our theory can be understood as a subgroup of this $E_{6(6)}$, which remains as an exact global symmetry. 

The connection between the $SU(4)$ gauge bosons in the $AdS$ theory and the 15 abelian ones from the flat space limit can be understood directly as a consequence of the vanishing gauge coupling in \eq{key}: In a non-abelian theory where the gauge coupling is absorbed in the kinetic term, so that it is
\begin{equation} S_{\text{non-abelian}}=\frac12\int \text{tr}(\mathbf{F}\wedge *\mathbf{F}),\end{equation}
the gauge coupling appears in the fieldstrenghts,
\begin{equation} F^a=dA^a+ gf^{abc}A^bA^c,\end{equation}
so that in the $g\rightarrow0$ limit, the theory abelianizes, and the lagrangian just looks like that of dim$(G)$ abelian vector bosons. This is precisely what happens in the infinite $N$ limit described above. Another way of saying it is that $g\rightarrow0$ replaces the Lie group $G$ by the tangent space at the identity, and so the gauge group becomes $\mathbb{R}^{\text{dim}(G)}$. One might worry that the limit $g\rightarrow0$ does not exist; indeed, \cite{Pernici:1985ju,Gunaydin:1985cu} claim this is the case; however, as explained in \cite{deWit:2002vt}, this is an artifact of a particular parameterization. Quoting from this reference: ``{The appearance of the inverse coupling constant $g^{-1}$ [...] in this term shows that, after gauge fixing, the theory no longer possesses a smooth limit to the ungauged theory. This phenomenon has been observed in the original construction of the SO(p,q) gauged theories [...] Note that the full Lagrangian [...] in contrast allows a smooth limit $g\rightarrow0$.}''

From whichever perspective, the conclusion is that we have arrived at a flat space quantum gravity with a global symmetry, which is in the Swampland, and therefore the theory we started with must be as well. The fact that the infinite distance limit group is $\mathbb{R}^{\text{dim}(G)}$ means that there are also 1-form symmetries \cite{Heidenreich:2020pkc}, and these imply that there cannot be any magnetically charged states under the abelian factors. This in turn runs afoul of the ordinary Distance \cite{Ooguri:2006in} and lattice Weak Gravity Conjectures, \cite{Montero:2016tif,Heidenreich:2016aqi}, which both demand the existence of these states in the infinite distance limit. As mentioned in the Introduction, see \cite{Cribiori:2022trc} for a connection with the lattice Weak Gravity Conjecture.

One could entertain the possibility that there is some other tower of states becoming light in the infinite $N$ limit, different from the usual tower of $S^5$ KK modes, which drives the cutoff of the five-dimensional theory to zero quickly enough, thereby avoiding the paradox; for instance, if one formally  replaces the $S^5$ by $S^5/\mathbb{Z}_n$, the tower of KK modes starts at ``level $n$'' relative to the unnormalized KK tower (although in this example one also loses supersymmetry; we are envisioning some other putative supersymmetric modification of the theory). Another possibility would be to imagine a theory like Vasiliev, which has a tower of light modes of arbitrarily high spin, and in which therefore sub-AdS locality loses meaning. In both these cases, and also in whatever else one does, however, some tower becoming light in five-dimensional Planck units is necessary to avoid a global symmetry in the infinite distance limit.  In $AdS$ units, this means
\begin{equation} m_{\text{Tower}}\ell \sim N^\alpha,\quad \alpha>\frac23.\end{equation}

\section{Examples where extreme scale separation is excluded}\label{sec:list}
The argument carried out above extends almost verbatim to any $AdS$ quantum gravity where the R-symmetry is gauged, because no flat-space limit of string theory has a gauged R-symmetry. This is a large classes of $AdS$ quantum gravities, since the R-symmetry current is often part of the stress-energy tensor multiplet. We can make a similar argument to the one above, using that the $d$-dimensional Planck length $\ell_d$ and gauge coupling are related to the stress-energy and current two-point functions as
\begin{equation} C_T\sim\left(\frac{\ell}{\ell_d}\right)^{d-2},\quad C_J\sim C_T\sim\frac{\ell^{d-4}}{g^2}, \end{equation}
where we have used that the R-symmetry currents are in the stress-energy tensor superconformal multiplet and thus its two-point function scales as $C_T$. This immediately leads to the conclusion that the gauge coupling satisfies
\begin{equation}\frac{g^2}{\ell_d^{d-4}}\sim C_T^{-\frac{2}{d-2}},\end{equation}
and so it goes to zero for any $d>2$ when the central charge is made large. With a list of all superconformal algebras in hand \cite{Cordova:2016emh}, one can compile Table \ref{t2}. We see that pure supergravity is excluded for most dimensions and supersymmetry algebras, with one exceptions: 4d $\mathcal{N}=1$, where the R-symmetry is trivial and no argument can be made. This is the case corresponding to DGKT and other putative candidates of scale-separated $AdS$ vacua, and for good reason: the absence of R-symmetry means precisely that there is no large extra dimension.

We also need a bit of extra care in the 4d $\mathcal{N}=2$ and 5d $\mathcal{N}=1$ cases, where the R-symmetry is $U(1)$. In the other cases, where the R-symmetry group is non-abelian, the gauge coupling is uniquely determined by the two-point function $C_J$, which is tied by supersymmetry to $C_T$. The normalization of non-abelian currents is uniquely fixed by their current algebra.  But this is not so in the abelian case, where the current can be rescaled by an arbitrary factor. A canonical nomalization is such that the charge of the operator of smallest charge is exactly 1, but this may not coincide with the normalization coming from the $C_J$ of the current appearing in the gravity multiplet; equivalently, the gravitino R-charge may be a large multiple (perhaps dependent on $C_T$ of the elementary charge). The R-charge of the gravitino cannot diverge in the large $C_T$ limit, where the R-symmetry is ungauged, leading to the same conclusion as in previous cases.  

\begin{table}[!htb]
\begin{equation*}
\begin{array}{|c|c|c|}
\hline\text{Dimension}&\text{\# of Q's}&\text{R-symmetry}\\\hline
 \multirow{4}{*}{$d=4$}&2\, (\mathcal{N}=1)\cellcolor{undecided}&--\cellcolor{undecided}\\
 &4\, (\mathcal{N}=2)\cellcolor{swamp}&\mathfrak{so}(2)\cellcolor{swamp}\\
  &6\, (\mathcal{N}=3)\cellcolor{swamp}&\mathfrak{so}(3)\cellcolor{swamp}\\
    &8\, (\mathcal{N}=4)\cellcolor{swamp}&\mathfrak{so}(4)\cellcolor{swamp}\\
     &10\, (\mathcal{N}=5)\cellcolor{swamp}&\mathfrak{so}(5)\cellcolor{swamp}\\
  &12\, (\mathcal{N}=6)\cellcolor{swamp}&\mathfrak{so}(6)\cellcolor{swamp}\\
       &16\, (\mathcal{N}=8)\cellcolor{swamp}&\mathfrak{so}(8)\cellcolor{swamp}\\\hline
  \end{array}
\quad\quad 
\begin{array}{|c|c|c|}
\hline \text{Dimension}&\text{\# of Q's}&\text{R-symmetry}\\\hline
\multirow{4}{*}{$d=5$}&4\, (\mathcal{N}=1)\cellcolor{swamp}&\mathfrak{u}(1)\cellcolor{swamp}\\
 &8\, (\mathcal{N}=2)\cellcolor{swamp}&\mathfrak{su}(2)\oplus\mathfrak{u}(1)\cellcolor{swamp}\\
  &12\, (\mathcal{N}=3)\cellcolor{swamp}&\mathfrak{su}(3)\oplus\mathfrak{u}(1)\cellcolor{swamp}\\
    &16\, (\mathcal{N}=4)\cellcolor{swamp}&\mathfrak{su}(4)\cellcolor{swamp}\\\hline
 d=6&8\, (\mathcal{N}=1)\cellcolor{swamp}&\mathfrak{su}(2)\cellcolor{swamp}\\\hline
  \multirow{2}{*}{$d=7$}&8\, (\mathcal{N}=(1,0))\cellcolor{swamp}&\mathfrak{su}(2)\cellcolor{swamp}\\
 &16\, (\mathcal{N}=(2,0))\cellcolor{swamp}&\mathfrak{sp}(2)\cellcolor{swamp}\\\hline
\end{array}
\end{equation*}
\caption{A table of all superconformal algebras in $d-1$ spacetime dimensions, taken from \cite{Cordova:2016emh}, corresponding to quantum gravities in $AdS_{d}$. We have shaded in red theories for which the arguments in the main text rule out the possibility of extreme scale separation; there is a single case shaded in yellow, where the arguments in this note do not work, but we do not know of an embedding in quantum gravity, either.}
\label{t2}
\end{table}


\section{Concluding remarks}\label{sec:conclu}

A common trope of the Swampland Program is the idea that the tower of states of the Distance Conjecture often appears to protect the theory from developing a global symmetry in the infinite distance limit \cite{Gendler:2020dfp}. In this short note we have pointed out how this idea can be straightforwardly applied in the $AdS$ context to rule out an extreme version of scale-separated examples that may look completely natural, and even appealing, from supergravity and CFT/boostrap arguments.  The maximal supersymmetry examples we have discussed are an ideal laboratory to confront the predictions of the Swampland program with CFT/bootstrap expectations due to the simplicity of the setup.

Our observations suggest that, from the CFT point of view, the pathology in these theories may be apparent in the bulk point limit of stress-energy supermultiplet correlation functions. Progress in this direction would be very exciting, since the appearance of a pathology would lead to a nice interpretation of Swampland conditions from a bootstrap perspective. 


Our simple arguments here only apply to extreme scale separation, and cannot be used in other examples such as DGKT. Extending them to these cases would involve understanding the precise relation between a small gauge coupling and the cutoff of the effective field theory in quantum gravity. Recent work in this direction includes \cite{Daus:2020vtf,Cordova:2022rer}. 

We finish this short note by noting that the theories we have ruled out are just supersymmetric versions of pure gravity in $AdS$. Things are usually easier with more supersymmetry, so our result are not encouraging for the $\mathcal{N}=0$ version of this question, where there is not even Bose-Fermi cancellation to ensure a small cosmological constant\footnote{For example for the holomorphic version of non-supersymmetric $AdS$$_3$ the evidence for the would be holographic dual is now pointing to the direction that most of them cannot exist \cite{Lin:2021bcp}.}. This paper serves as a cautionary note, that one should not take on face value a low energy description of $AdS$ and holography and draw conclusions based on it, without showing the existence of its UV completion. The experience with the Swampland principles lead us to believe that most consistent looking low energy theories are indeed in the Swampland.\\

\textbf{Acknowledgements:}
We thank Shai Chester, Daniel Jafferis, Severin L{\"u}st, Leonardo Rastelli, and Thomas Van Riet  for interesting discussions and comments. The work of MM and CV is supported by a grant from the Simons Foundation (602883,CV) and by the NSF grant PHY-2013858. MR would like to thank Cumrun Vafa and the Harvard Theory group for hospitality, and the generosity of the Bershadsky Fellowship for partial support.  The work of MR is partially supported by NSF grant PHY-2210533.

\bibliographystyle{JHEP}

\bibliography{refs}

\end{document}